\documentclass[11pt]{article}
\relax
\catcode`\@=11
\newcommand{\la}[1]{\label{#1}}
\newcommand{\ii}{{\rm i}}                               
\newcommand{\dd}{{\rm d}}                                
\newcommand{\vder}[2]{{\frac{\delta #1}{\delta #2}}}        
\newcommand{\half}{{1 \over 2}}                           
\newcommand{\Lie}{{\cal L}}

\newcommand{\DD}{{\cal D}}
\newcommand{\doo}{{\partial}}
\newcommand{\sgn}{{\rm sign \  } }
\newcommand{\ind}{{\rm ind \  }  }

\newcommand{\Sp}{{\rm Sp  }  }

\newcommand{\Abs}[1]{{\biggl\vert {#1} \biggr\vert}}
\newcommand{\tr}{{\rm  tr\ }}

\newcommand{\Det}{{\rm Det }}

\newcommand{\slQ}{\slash\hskip-0.24cm {Q}}
\newlength{\ypit}

\newcommand{\be}{\begin{equation}}
\newcommand{\ee}{\end{equation}}
\newcommand{\ba}{\begin{eqnarray}}
\newcommand{\ea}{\end{eqnarray}}

\newcommand{\bea}{\begin{eqnarray}}
\newcommand{\eea}{\end{eqnarray}}

\newcommand{\comm}[2]{ \left[ #1 ,  #2 \right] }

\newcommand{\Ker}{{\rm Ker}\:}

\newcommand{\bastar}{\begin{eqnarray*}}
\newcommand{\eastar}{\end{eqnarray*}}

\newcommand{\poisson}[2]{\{ #1 , #2 \}}

\newcommand{\X}{{\cal X}}

\begin{document}

\begin{titlepage}
\begin{flushright}
\end{flushright}

\vskip 0.5truecm

\begin{center}
{ \Large \bf
On The Arnold Conjecture And The \\
Atiyah-Patodi-Singer Index Theorem
}
\end{center}

\vskip 1.0cm

\begin{center}
{\bf Mauri Miettinen$^{\star}$ $~$ 
and $~$ Antti J. Niemi$^{\#}$ }  \\
\vskip 1.0 cm 
{\it $^{\star \#}$Department of Theoretical Physics, 
Uppsala University,
P.O. Box 803, S-75108, Uppsala, Sweden \\
$^{\star \#}$The Mittag-Leffler Institute, 
Aurav\"agen 17, S-182 62 Djursholm, Sweden \\ 
$^{\#}$Helsinki Institute of Physics, P.O. Box 9, FIN-00014 
Helsinki University, Finland \\
}
\vskip 1.5cm
\end{center}

\noindent
The Arnold conjecture yields a lower bound to the
number of periodic classical trajectories in a Hamiltonian
system. Here we count these trajectories
with the help of a path integral, which we inspect
using properties of the spectral flow of 
a Dirac operator in the background of a $\Sp(2N)$ 
valued gauge field. We compute the spectral flow from
the Atiyah-Patodi-Singer index theorem, and apply the
results to evaluate the path integral using localization
methods. In this manner we find a lower bound to the 
number of periodic classical trajectories which is consistent 
with the Arnold conjecture.
\vskip 0.5cm

\rm
\noindent

\vfill

\begin{flushleft}
\rule{5.1 in}{.007 in}\\
$^{\#}${{ \bf
niemi@teorfys.uu.se } $~~~~$ \\
 }
\end{flushleft}

\end{titlepage}

Periodic classical trajectories are important in a large number
of problems.  Examples range from classical
integrable models where the motion is a periodic flow on 
invariant torii, to quantum chaos where 
the energy spectrum is computed 
by the Gutzwiller (Selberg) trace formula. For non-integrable 
models exact analytical results are sparse, and in 
general even a classification of trajectories 
becomes arduous. Some 35 years ago Arnold conjectured \cite{arnold} 
that in an arbitrary Hamiltonian system the number 
of periodic classical solutions is bounded 
from below by the number of critical 
points of a function. If the Hamiltonian is energy 
conserving {\it i.e.} has no explicit time 
dependence, this follows immediately 
from conventional Morse inequalities since every 
critical point of the Hamiltonian is 
trivially a periodic classical solution. But 
in the general case when the Hamiltonian is explicitly time 
dependent so that energy is not conserved, the conjecture has 
only been verified in special cases. The modern approach to 
the Arnold conjecture is due to Floer \cite{floer} who formulated
it using infinite dimensional Morse theory. 
His approach has led to extensive investigations 
\cite{floer2} and
applications in seemingly disparate subjects, 
most notably topological quantum field theories 
and strings \cite{witten}. 

Here we shall study relations between the Arnold conjecture 
and canonical path integrals. We suggest an intuitive 
approach for proving the conjecture using path 
integral localization techniques. Our arguments are 
based on methods of topological quantum field theories and
properties of spectral flow in a 1+1 dimensional Dirac operator 
in the background of an adiabatically varying $\Sp(2N)$ valued 
classical gauge field. For this we consider Hamilton's equation of 
motion on a $2N$-dimensional 
phase space {\it i.e.} a symplectic manifold $\Gamma$ with 
local coordinates $x^i$,  
\be
\dot{x}^i \ = \ \omega^{ij} \doo_j H(x,t) \ \equiv \ \X^i_H (x,t) 
\la{eqm} 
\ee 
Here $H(x,t)$ is in general an explicitly time dependent Hamiltonian 
and $\X_H (x,t) $ is the corresponding Hamiltonian vector field. The 
antisymmetric $\omega^{ij} = \poisson{x^i}{x^j}$ is the matrix inverse
of the symplectic two-form $\omega = \half \omega_{ij} \dd x^i \wedge 
\dd x^j$ on $\Gamma$. The solutions of (\ref{eqm}) are critical
points of the classical action
\be
S \ = \ \int \dd t \left[ \vartheta_i \dot{x}^i - 
H(x,t) \right]
\la{act}
\ee
where $\vartheta$ is the symplectic potential, $\omega_{ij} = 
\partial_i \vartheta_j -\partial_j \vartheta_i$. 
For a $T$-periodic Hamiltonian $H(x,t) = H(x, t+T)$ 
defined on a compact $\Gamma$ the Arnold conjecture 
states that the number of non-degenerate and contractible 
$T$-periodic  solutions is bounded from 
below by the sum of the Betti numbers $B_p$ on $\Gamma$, 
\be
\# \{ {\rm T-periodic ~ solutions } \} \
\geq \ \sum B_p
\la{arn}
\ee 
When the $t$-dependence of $H$ is trivial (\ref{arn}) 
reduces to the conventional finite 
dimensional Morse inequality \cite{morse}
since every critical point of $H$ is trivially 
$T$-periodic. But for a generic $t$-dependent 
Hamiltonian (\ref{arn}) 
remains unproven. 

The number of $T$-periodic solutions 
is counted by the path integral
\be
Z \ = \ \sum_{\delta S =0} 1
\ = \ \int \DD x \ \delta \left[ \vder{S}{x} \right]
\cdot \Abs{\ \Det \left[ {\delta^2 S \over \delta x^2 }   \right] }
\la{path1}
\ee
where we have the determinant of the Hessian
\be
{\delta^2 S \over \delta x^i
(t_1) \delta x^j (t_2) } \ =  
\ \left\{ \omega_{ij}\partial_{t_1} \ + \  
\frac{1}{2}\left( \partial_j \omega_{ik} + \partial_i
\omega_{jk} \right) \dot x^k - \partial_i \partial_j H 
\right\} \delta(t_1 - t_2)
\la{hes}
\ee
which is a symmetric first-order differential operator.
The zero-modes of (\ref{hes}) are Jacobi fields, but
we assume that the $T$-periodic solutions of (\ref{eqm}) are 
isolated and non-degenerate so that 
the Jacobi fields are absent. 
The difficulty in establishing (\ref{arn}) is 
then due to the absolute signs in (\ref{path1}),
alternatively due to the first-order nature of the Hessian: 
The Morse index of (\ref{act}) coincides with the number of negative 
eigenvalues of (\ref{hes}). But since the spectrum of (\ref{hes}) 
is unbounded from below this number is infinite and needs 
to be regulated appropriately.  

We first consider (\ref{path1}) without the absolute value.
Formally, the path integral then evaluates to the Euler 
characteristic on the loop space $\rm L\Gamma$,
\be
Z \ = \ \sum_{\delta S=0} (-)^k \ = \ \int\DD x  \ 
\delta \left[ \vder{S}{x} \right]
\Det \left[  { \delta^2 S \over \delta x^2 }   \right]
\la{path2}
\ee
where $k = 0$ or $1$ depending on whether the Morse index of the
pertinent critical point of $S$ is even or odd, {\it i.e.} whether 
the Hessian has an even or odd number of negative 
eigenvalues in some proper regularization scheme. 
We introduce a commuting Lagrange multiplier $\lambda$ 
and anticommuting ghosts $\psi, \bar{\psi}$, and re-write (\ref{path2}) as
\ba
Z &=& \int \DD \Phi
\exp \left\{ \ii \int_0^T \dd t \left[ 
\lambda_i(\dot{x}^i -\X_H^i ) +
\bar{\psi}_i (\delta_j^i \doo_t + 
\doo_j \X_H^i )\psi^j  \right] 
\right\} \cr 
& = & \int \DD \Phi \exp
\left\{ \ii \int_0^T \dd t \ \left[ \lambda_i
\dot{x}^i + \bar{\psi}_i \dot{\psi}^i
- \lambda_i \X^i_H - \psi^i 
\doo_i \X^j_H \bar{\psi}_j  \right] \right\} 
\la{fpi} 
\ea 
where we denote by $\Phi$ collectively
the $T$-periodic variables $(x, \lambda, \psi, \bar{\psi})$.
This has the structure of a canonical path integral 
in a graded phase space in Darboux
coordinates with Poisson brackets
\be 
\poisson{ \lambda_i}{x^j} \ = \
\poisson{\bar{\psi}_i}{\psi^j} \ = \ \delta_i^j 
\la{pb}
\ee
We note that the Lagrange multiplier $\lambda$ provides  
a realization of derivation {\it w.r.t.} the variable
$x$. Similarly we identify $\psi \sim dx$, the basis one-forms, 
and $\bar\psi$ is the corresponding basis of contractions. 
We define the exterior derivative in 
the loop space $\rm L\Gamma$
\be
\dd \ = \ \int_0^T \dd t~ \psi^i \lambda_i 
\la{loop}
\ee 
We may view $\dd$ as a nilpotent BRST operator
and the action $S_{\rm ext}$ in (\ref{fpi}) is  
a BRST commutator,
\be
S_{\rm ext} \ = \ \lambda_i
\dot{x}^i + \bar{\psi}_i \dot{\psi}^i
- \lambda_i \X^i_H - \psi^i 
\doo_i \X^j_H \bar{\psi}_j \ = \   \comm{\dd}{ \Psi } \ 
= \ \comm{\dd}{ \ (\dot{x}^i 
- \X^i ) \bar{\psi}_i } \ = \ 
\Lie_S
\la{tft}
\ee 
with $\Lie_S$ the Lie derivative along the loop space vector 
field $\X_S = \dot{x} - \X_H$. These relations mean that
the path integral (\ref{fpi}) determines 
a topological quantum theory \cite{blau}. 

The Fradkin-Vilkovisky theorem \cite{BV} states that 
(\ref{fpi}) is invariant under an infinitesimal 
variation $\Psi \to \Psi + \delta\Psi$ of the gauge
fermion in (\ref{tft}). 
An arbitrary infinitesimal coordinate 
transformation $x \rightarrow \tilde{x} = x+ \epsilon (x)$ in 
the original phase space $\Gamma$ is an example of 
such a variation, with 
\[
\Psi \ \to \ \Psi \ + \ \frac{\partial \epsilon^i}{\partial x^j}
({\dot x}^j - \X_H^j ) \bar\psi_i
\]
provided we also conjugate the BRST operator according to
\[
\dd \ \to \ e^{- \lambda_i \epsilon^i} \ \dd \ e^{ \lambda_i 
\epsilon^i }
\]
Since this conjugation does not alter the 
cohomology of $\dd$ it has no
effect on the path integral, which is left intact. 
As a consequence we can 
inspect the local structure of the path integral by 
selecting (local) Darboux coordinates on $\Gamma$, 
\be
\omega \ \to \ \omega_0 \ = \ \sum_{k=1}^N d p_k \wedge d q^k
\la{dar}
\ee 
In these coordinates 
\be
A_{~ i}^j \ = \ \doo_i \X^j_H \ = \
\omega^{jk}_0 \doo_i\doo_k H
\la{gauge}
\ee 
is a $\Sp(2N)$-valued traceless product of the antisymmetric 
$\omega_0^{-1}$ and the symmetric Hessian $\doo^2 H$, and  
the ghost part in ({\ref{fpi}) simplifies into  
\be
S_{\rm F} = \int_0^T \dd t \
\bar{\psi} (\doo_t \ + A ) \psi
\la{sf}
\ee
This is reminiscent of a (0+1) dimensional Dirac Lagrangian
in the background of a gauge field. Indeed, the Darboux 
coordinates are defined only modulo canonical 
transformations corresponding to arbitrary
$t$-dependent conjugations of the symplectic 
two-form (\ref{dar}) by elements of $\Sp(2N)$,
\[
\omega_0 \ = \ U^{-1}(t) \omega_0 U(t)
\]
Under these $\Sp(2N)$ conjugations the anticommuting 
ghosts transform according to  
\ba
\psi(t) \ & \rightarrow & \  U(t) \psi(t) \cr
\bar{\psi}(t) \ &  \rightarrow & \ \bar{\psi}(t) U^{-1}(t)
\ea
and (\ref{sf}) remains invariant if $A_{~ i}^j(t)$ 
transforms like a $\Sp (2N)$-valued gauge field 
\be
A(t) \rightarrow U(t) A(t) U^{-1}(t) + {U}(t) 
\dot{U}^{-1}(t)
\la{gtr}
\ee

Following \cite{eli} we introduce
\[
\Lambda (t) \ = \ {\rm T} \exp \left( \ \int\limits_{0}^t
A(t') dt' \ \right)
\]
and define a $\Sp(2N)$ gauge transformation with
\be
U(t) \ = \ e^{-A_0 t} \cdot \Lambda^{\dagger} (t)
\la{tran1}
\ee
where $A_0$ is defined by
\be
\Lambda(T) \ = \ \exp( - A_0 \cdot T)
\la{tran2}
\ee
which is essentially the average of $A(t)$ over the 
period $T$. This gauge transformation maps 
our $\Sp (2N)$-valued gauge field $A(t)$ into 
the $t$-independent average $A_0$,
\be
U(t) A(t) U^{-1}(t) + {U}(t) \dot{U}^{-1}(t) \ = \ A_0
\la{tran3}
\ee
As a consequence the ghost Lagrangian (\ref{sf})
depends only on the $t$-independent $A_0$. 
Moreover, if we decompose the $T$-periodic 
Hamiltonian $H(x,t)$ into its Fourier modes
\be
H(x,t) \ = \ H_0(x) \ + \ \sum_{n\not= 0} 
H_n(x) \cdot \exp(\frac{2 
\pi \ii n t}{T}) 
\la{fourier} 
\ee 
where the $H_n(x)$ have no explicit $t$-dependence
and $H_0(x)$ is the average of $H(x,t)$ over the period $T$, 
we conclude that the gauge transformation (\ref{tran3}) 
replaces the original Hamiltonian $H(x,t)$ by its $t$-independent 
average $H_0(x)$ in the ghost Lagrangian (\ref{sf}).

In the following we shall assume 
that the average $H_0(x)$ is a Morse function
with only isolated and 
non-degenerate critical points. 

Let $\Psi$ denote the original gauge
fermion in (\ref{tft}) and let $\Psi_0$ denote
the corresponding
gauge fermion with $H(x,t)$ replaced by the 
average $H_0(x)$. We define the infinitesimal 
change of variables  
\be
\Phi \ \to \ \Phi \ + \ \epsilon \cdot \{ \dd , \Phi \}
\cdot \int\limits_0^T ( \Psi - \Psi_0)
\la{local}
\ee  
with $\epsilon$ an infinitesimal parameter. We find that
in (\ref{fpi}), (\ref{tft}) the {\it only} effect 
of (\ref{local}) is the shift
\[
\Psi \ \to \ (1-\epsilon) \Psi \ + \ \epsilon \Psi_0
\]
and by repeating this change of variables we can replace 
$H(x,t)$ by $H_0(x)$ in (\ref{fpi}). Hence (\ref{fpi}) 
depends only on the $t$-independent average $H_0(x)$ of
the Hamiltonian.

The Hessian (\ref{hes}) has an eigenvalue 
problem
\be
{D} f_n (t) =  \lambda_n f_n(t) \ \ \; , 
\ \ \ \ \ ~~~f_n(0)=f_n(T)
\ee
where the eigenvalues $\lambda_n$ are invariant under
$\Sp(2N)$ gauge transformations that are continuously 
connected to the identity. Since we assume that 
there are no Jacobi fields, 
in the background of a classical solution all $\lambda_n \not=0$.
If we introduce the $\eta$-invariant of the Hessian
\be
\eta_{{D}} \ = \ \lim_{s \rightarrow 0} \sum_{\lambda_n\not=0}
\sgn (\lambda_n ) \ | \lambda_n |^{-s} 
\la{eta} 
\ee 
we can remove the absolute 
value of the determinant 
in (\ref{path1}) by 
\be
\Abs{\ \Det [{D}] \ } \ \to \ 
\Det [{D}] \cdot \exp \left({\ii \pi \over 2} 
\eta_{{D} } \right) 
\la{releta0}
\ee
This implies 
that the number of $T$-periodic classical trajectories
can be counted by the following generalization of (\ref{fpi}),
\be
{\cal Z} \ = \ 
\int \DD \Phi \exp \left\{  \ii
\int_0^T \dd t \ \left[ \lambda_k \dot{x}^k + 
\bar{\psi}_k \dot{\psi} - \lambda_k
\X^k_H - \psi^k \doo_k \X^l_H \bar{\psi}_l \right]
+ {\ii \pi \over 2} \eta_{{D}}  \right\}
\la{fpi2}
\ee
The path integral (\ref{fpi}) depends only on the $t$-independent
average $H_0(x)$. In particular, in the ghost action (\ref{sf}) 
the Hamiltonian $H(x,t)$
can be converted into $H_0(x)$ by a $\Sp(2N)$ gauge 
transformation which leaves the $\eta$-invariant of the Hessian
intact. As a consequence we conclude that the path integral
(\ref{fpi2}) should not change if 
we replace $H(x,t) \to H_0(x)$ in it.

The Atiyah-Patodi-Singer index theorem \cite{APS} relates
the $\eta$-invariant of a (0+1) 
dimensional Dirac 
operator to the index of a (1+1) dimensional Dirac 
operator. Accordingly we extend our Hessian (\ref{hes}) 
to the following two-dimensional Dirac 
operator on a cylinder $S^1 \times {\bf R}$ 
\be
\slQ \ = \ \pmatrix{ 0 & \doo_s + {D}_s \cr - \doo_s + 
{D}_s & 0} \ = \
i \sigma_2 \otimes \doo_s +  \sigma_1 \otimes {D}_s  \ = \
\pmatrix{0 & Q^\dagger \cr Q & 0
}
\la{dirac2}
\ee
Here ${D}_s$ with $s \in {\bf R}$ is a one-parameter 
family of symmetric Hessians on  
the cylinder. We define them 
by promoting our $T$-periodic solution $x(t)$ into a 
one-parameter family of $T$-periodic trajectories $u(s,t)$, 
an instanton that adiabatically interpolates 
between $x(t)$ at $s \to + \infty$
and another $T$-periodic $y(t)$ at $s\to - \infty$,
\ba
\lim_{s \rightarrow + \infty}u(s,t) &=& 
x(t) 
\cr 
\lim_{s \rightarrow - \infty} u(s, t)&=& y(t)
\la{tun}
\ea 
The operators ${D}_s$ have an $s$-dependent eigenvalue problem 
\be 
{D}_s f_n (t, s) =  \lambda_n (s)
f_n (t,s) \ \ , \ \ \ \ \ \ \ \ f_n(0,s) = f_n (T,s) 
\la{2dop}
\ee 
and in the absence of Jacobi fields at $s \to \pm \infty$
all $\lambda_n(\pm \infty) \not= 0$.
But as a function of the adiabatic parameter $s$, the 
operators $D_s$ in general exhibit a non-trivial 
spectral flow with some 
of the eigenvalues $\lambda_n (s)$ crossing zero when 
the parameter $s$ evolves. This spectral flow can be computed
by applying the Atiyah-Patodi-Singer index theorem to the Dirac 
operator $\slQ$. It relates the index of $\slQ$ on the cylinder 
to the $\eta$-invariants of the operator ${D}_s$ at
the $s\to\pm\infty$ ends of the cylinder,
\ba
\ind \slQ \ & = & \ {\rm dim \ Ker}Q \ - 
\ {\rm dim \ Ker}Q^\dagger \cr
& = & \ \frac{1}{2} \left( \ \eta_{{D}} [y(t)] -\eta_{{D} }
[x(t)] \ \right) \ + \ {1 \over 2\pi} \int_{S^1 \times {\bf
R}} \tr  F 
\la{aps}
\ea 
Here $F = \doo_s (\omega_0 A)$ is a $\Sp(2N)$-valued 
field strength: We have a two dimensional $\Sp(2N)$ gauge 
field $A$ with components $A_t = \omega_0 A$ and $A_s =0 $. 
In general the contribution from $F$ cancels 
the continuous part in the difference of the $\eta$-invariants 
in (\ref{aps}). But since we have verified that 
in the Darboux variables (\ref{dar}) the $\Sp(2N)$ 
valued gauge field $\omega_0 A$ is explicitly traceless,
the relative $\eta$-invariant in (\ref{aps})
\be
\Delta \eta_{D} \ = \  \eta_{{D}} [y(t)] \ - \  \eta_{{D}} [x(t)]
\la{releta1}
\ee
should be an integer, it coincides with the net spectral 
flow between the two configurations $x(t)$ and $y(t)$,
\be
\Delta\eta_D \ \sim \ {\rm spectral \ flow \ of} \ \slQ
\la{releta2}
\ee
In particular (\ref{releta1}) determines the relative sign 
between the determinants of the Hessians for the two 
trajectories $y(t)$ and $x(t)$. Hence $\Delta\eta_D$
coincides with the relative Morse index between the 
two trajectories, which is also the difference between their 
Maslov indices \cite{floer2}. 

Instead of (\ref{fpi2}) we proceed with the more general
\be
{\cal Z} = \int \DD \Phi \exp \left[ \ii
\int_0^T \dd t \ \left( \lambda_k \dot{x}^k + 
\bar{\psi}_k \dot{\psi} - \lambda_k
\X^k_H - \psi^k \doo_k \X^l_H \bar{\psi}_l  
\right) + {\ii \pi \over 2} \Delta \eta_{{D}}  \right] \;.
\la{fpi3}
\ee
with $\Delta \eta_{D}$ 
the relative $\eta$-invariant
{\it w.r.t.} a fixed $T$-periodic 
reference trajectory $y(t)$. It is integer valued, and
twice the index of $\slQ$ in the background of the
pertinent instanton $u(s,t)$. Since (\ref{fpi2}) remains
intact when we replace $H(x,t)$ by its average $H_0(x)$,
we can expect that (\ref{fpi3}) also remains intact 
when we set $H(x,t) \to H_0(x)$.  

Since the relative $\eta$-invariant in (\ref{fpi3})
coincides with the integer valued index of the Dirac
operator $\slQ$, it should not change under a continuous
change of variables. This suggests that we can evaluate
the integral (\ref{fpi3}) by localization.
Accordingly we decompose the variables $\Phi$ 
in their $t$-independent and $t$-dependent components, 
$\Phi \to \Phi_0 + \Phi_t$, and set 
$x(t) \to x_0 + \alpha x_t$, $\psi (t) \to \psi_0 
+ \sqrt{\alpha} \psi_t$, and $\bar{\psi}(t) \to \bar{\psi}_0 
+ \sqrt{\alpha} \bar{\psi}_t$ with $\alpha$ a 
parameter. The corresponding super-Jacobian 
in (\ref{fpi3}) is trivial, and we localize 
by setting $\alpha \to 0$. In this limit 
the path integral (\ref{fpi}) is known to localize to 
the Poincar\'e-Hopf representation of the Euler 
characteristic of the 
phase space $\Gamma$ \cite{nie},
\be 
Z \ = \ \sum_{\X_{H_0}=0} \sgn\det (\doo \X_{H_0} ) \ = \
\sum_{k=0}^{2N} (-1)^k m_k \ = \ \sum_{k=0}^{2N} (-1)^k B_k \ = \
\chi( \Gamma)
\ee 
where $m_k$ denotes the number of critical points with Morse 
index $k$. In a similar manner we expect 
the path integral (\ref{fpi3}) to localize into
\be 
{\cal Z} \ = \ \sum_{\X_{H_0}=0}
\sgn \det (\doo \X_{H_0} ) \exp \left( + {\ii \pi \over 2} \Delta 
\eta_{{D} } \right)
\la{final1}
\ee 
where $\Delta \eta_D$ is now the relative $\eta$-invariant 
for the critical points of $H_0(x)$ {\it w.r.t.} the 
fixed periodic reference trajectory $y(t)$. We 
select this reference trajectory 
so that it coincides with a critical point 
of $H_0(x)$ with Morse index $0$. According to 
the standard finite dimensional Morse 
inequality $m_p \geq B_p$ and $B_0 \geq 1$ which ensures that
such a critical point always exists. With this,  
the pertinent Dirac index (\ref{aps}) simplifies into
\be
\ind \slQ \ \ = \ \ \dim \Ker Q \ -\ \dim Ker Q^\dagger \ 
= \ \dim \Ker Q
\la{trunc}
\ee
This follows directly from the zero mode equation in the adiabatic
limit, where we can use the Ansatz 
\[
g_n (s,t) \ = \ \phi_n (s) \rho_n (t)
\]
with
\[
D_s \rho_n (t) \ = \ \lambda_n (s) \rho_n (t)
\]
This reduces the zero mode equations for 
$Q$ and $Q^{\dagger}$ to
\[
(\doo_s \pm \lambda_n(s) ) \phi_n (s) \ = \ 0
\]
with solutions
\[
\phi_n (s) \ = \ \exp \left[\mp \int_0^s \dd s' 
\lambda_n (s' )   \right] \phi_n (0)
\]
respectively. Hence $Q^\dagger = \doo_s +D_s$ does not have 
a normalizable zero mode and (\ref{trunc}) follows. 
In particular, the spectral flow is only from positive eigenvalues 
of the Hessian to its negative eigenvalues, which 
implies that the contribution from the relative 
$\eta$-invariant $\Delta \eta_D$ in (\ref{final1}) 
coincides with the Morse index $m_{H_0} (x) = k$ of the critical 
point. Since $\sgn \det (\doo \X_{H_0} )= (-1)^k$ and 
Morse inequalities yield $m_k \geq B_k$, we then conclude
that modulo an overall, irrelevant sign the path integral 
(\ref{fpi3}) localizes into 
\[
Z \ = \ \sum_{\X_{H_0} = 0} \sgn \det (\doo \X_{ H_0} ) 
\exp( - \ii \pi k ) \ = \ \sum_{\X_{H_0} =0} 1 \ = \ 
\sum_{k=0}^{2N} m_{k} \ \geq \ \sum_{k=0}^{2N} B_{k} 
\] 
As a consequence we have a lower bound estimate for the number 
of periodic classical trajectories which is
consistent with the Arnold 
conjecture.
\vskip 0.5cm
In conclusion, we have studied relations between a path integral and the
Arnold conjecture. We have related 
the path integral to the properties of a two dimensional Dirac 
operator and applied the Atiyah-Patodi-Singer index theorem
to evaluate it using localization methods. In this manner we 
have found a lower bound estimate for the number of periodic classical 
trajectories which is consistent with the Arnold conjecture. 
Our approach is based on intuitive path integral arguments,
and we hope that a rigorous derivation 
of the conjecture could be developed along these lines. 

\vskip 1.4cm
We thank Topi K\"arki for discussions. Our research has been 
partially supported by NFR Grant F-AA/FU 06821-308.
This article reports collaboration that was largely completed
by March 2nd 1999, when Mauri Miettinen unexpectedly passed away
just one month before his scheduled Thesis defence.

\end{document}